\documentclass[12pt,paper,a4paper]{JHEP3} % 10pt is ignored!

\JHEPspecialurl{http://jhep.sissa.it/JOURNAL/JHEP3.tar.gz}

\usepackage{epsfig,multicol,bbm}
\usepackage{amssymb}
\usepackage{cite}

\newcommand\fverb{\setbox\pippobox=\hbox\bgroup\verb}
\newcommand\fverbdo{\egroup\medskip\noindent%
                        \fbox{\unhbox\pippobox}\ }
\newcommand\fverbit{\egroup\item[\fbox{\unhbox\pippobox}]}
\newbox\pippobox
\newcommand{\beq}{\begin{equation}}
\newcommand{\eeq}{\end{equation}}
\newcommand{\bea}{\begin{eqnarray}}
\newcommand{\eea}{\end{eqnarray}}
\newcommand{\bem}{\begin{multline}}
\newcommand{\eem}{\end{multline}}
\newcommand{\beg}{\begin{gather}}
\newcommand{\eeg}{\end{gather}}

\def\eq#1{{Eq.~(\ref{#1})}}

\def\eq#1{{Eq.~(\ref{#1})}}

\newcommand{\as}{\alpha_s}

\title{Non-linear QCD meets data: A global analysis of lepton-proton
  scattering with running coupling BK evolution}

\author{Javier L.  Albacete$^1$, N\'estor Armesto$^2$, Jos\'e
  Guilherme Milhano$^3$ and Carlos A. Salgado$^2$

\vspace{0.1in}

{\it $^1$ European Center for Theoretical Studies in Nuclear Physics
  and Related Areas (ECT*), Strada delle Tabarelle 286, I-38050
  Villazzano (TN), Italy

 $^2$ Departamento de F\'{\i}sica de Part\'{\i}culas and
IGFAE,
Universidade de Santiago de Compostela,
 E-15706 Santiago de Compostela, Spain

$^3$ CENTRA, Instituto Superior T\'ecnico (IST),
Av. Rovisco Pais, P-1049-001 Lisboa, Portugal

\vskip 0.1in

E-mail addresses: {\tt albacete@ect.it, nestor.armesto@usc.es,
  guilherme.milhano@ist.utl.pt, carlos.salgado@usc.es}. } }

\preprint{ECT*-09-01}     % OR: \preprint{Aaaa/Mm/Yy\\Aaa-aa/Nnnnnn}
\abstract{ We perform a global fit to the structure function $F_2$
  measured in lepton-proton experiments at small values of
  Bjorken-$x$, $x\le 0.01$, for all experimentally available values of
  $Q^2$, $0.045 \ {\rm GeV}^2\le Q^2 \le 800 \ {\rm GeV}^2$. We show
  that the recent improvements resulting from the inclusion of running
  coupling corrections allow for a description of data in terms of
  non-linear QCD evolution equations. In this approach $F_2$ is
  calculated within the dipole model with all Bjorken-$x$ dependence
  described by the running coupling Balitsky-Kovchegov equation. Two
  different initial conditions for the evolution are used, both
  yielding good fits to data with $\chi^2/{\rm d.o.f.}<1.1$. Data
  for the proton longitudinal structure function $F_L$, not included
  in the fits, are also well described. Our analysis allows to
  perform a first principle extrapolation of the proton-dipole
  scattering amplitude. We provide predictions for
  $F_2$ and $F_L$ in the kinematical regions of interest for future
  colliders and ultra-high energy cosmic rays. A numerical implementation of our results down
  to $x=10^{-12}$ is released as a computer code for public use.}

\begin{document} 

%\setcounter{page}{1}

%%%%%%%%%%%%%%%%%%%%%%%%%%%%%%%%%%%%%%%%%%%%%%%%%%%%%%%%%%%%%%%%%%%%%%%%%%%%%
\section{Introduction} \label{intro}

The experimental data collected in electron-proton deep inelastic
scattering (DIS) experiments
\cite{Adams:1996gu,Arneodo:1996qe,Abt:1993cb,Ahmed:1995fd,Aid:1996au,Adloff:1997mf,Adloff:1999ah,Adloff:2000qk,Derrick:1993fta,Derrick:1994sz,Derrick:1995ef,
  Derrick:1996hn,Breitweg:1997hz,Breitweg:1998dz,Breitweg:2000yn,Chekanov:2001qu,:2008tx,Collaboration:2009na}
at small values of Bjorken-$x$ constitute one of the most valuable
sources of information to test and explore the high-energy limit of
Quantum Chromodynamics (QCD).  The standard analyses (see
\cite{Dittmar:2009ii} and references therein) of these data are usually
made in the framework of fixed order DGLAP evolution equations
%%%%Gui
% and resummations schemes have also been essayed
in which various resummation schemes have also been essayed.
%%%%
%%%%Gui
%On the other hand a description of available data in terms of the non-linear QCD evolution
%equations \cite{Jalilian-Marian:1997gr,Jalilian-Marian:1997dw,Kovner:2000pt,Weigert:2000gi,Iancu:2000hn,Ferreiro:2001qy,Balitsky:1996ub,Kovchegov:1999yj} has
%been elusive so far, although phenomenological analyses (see
%e.g. \cite{Golec-Biernat:1998js,Iancu:2003ge,Gotsman:2002yy,Albacete:2005ef,Kowalski:2003hm,Kowalski:2006hc,Goncalves:2006yt}) are most suggestive of the presence of {\it
%  saturation} effects, a crucial physical ingredient for the
%description of high-energy scattering, in the small-$x$ domain of DIS.
%%%%
On the other hand a description of available data in terms of the non-linear QCD evolution
equations \cite{Jalilian-Marian:1997gr,Jalilian-Marian:1997dw,Kovner:2000pt,Weigert:2000gi,Iancu:2000hn,Ferreiro:2001qy,Balitsky:1996ub,Kovchegov:1999yj} has --- despite phenomenological analyses (see e.g. \cite{Golec-Biernat:1998js,Iancu:2003ge,Gotsman:2002yy,Albacete:2005ef,Kowalski:2003hm,Kowalski:2006hc,Goncalves:2006yt}) being most suggestive of the presence of {\it
  saturation} effects, a crucial physical ingredient for the
description of high-energy scattering in the small-$x$ domain of DIS --- been elusive so far.

The saturation phenomenon is closely related to unitarity of the
quantum field theory and is characteristic of dense partonic
systems. It admits an intuitively clear physical picture in the
infinite momentum frame. There, the gluon distribution function
$xG(x,Q^2)$ can be interpreted as the number of gluons in the proton
wave function localized within a transverse area inversely
proportional to the photon virtuality $Q^2$, and carrying a fraction
of the proton longitudinal momentum $x$. For fixed $Q^2$, the number
of gluons in the proton wave function increases with decreasing $x$
due to additional gluon emission or gluon branching. Such growth of
gluon densities has been experimentally observed at HERA and, if
extrapolated towards smaller values of $x$, would threaten the
unitarity of the theory. Hence,  the proton gets denser and gluon-gluon
recombination processes, which are essentially non-linear, slow down
the non-abelian avalanche towards small-$x$. This mechanism tames the
subsequent growth of gluon densities, i.e. they {\it saturate}, thus
preventing unitarity violations. The intrinsic momentum scale that
determines the separation between the dilute and dense domains in the
proton wave function is the saturation scale $Q_s^2(x)$. This scale
can be understood as the inverse transverse area inside which the
probability of finding more than one gluon is of order one. It is a
dynamic scale whose growth is determined by the interplay between the
linear, radiative processes and the non-linear, recombination ones.

All these qualitative ideas are cast in a definite theoretical
framework, the Color Glass Condensate (CGC). The CGC is endowed with a
set of perturbative, non-linear evolution equations, the
Jalilian-Marian--Iancu--McLerran--Weigert--Leonidov--Kovner (JIMWLK)
equation \cite{Jalilian-Marian:1997gr, Jalilian-Marian:1997dw,
  Kovner:2000pt, Weigert:2000gi, Iancu:2000hn,Ferreiro:2001qy} and the
Balitsky-Kovchegov (BK) equation
\cite{Balitsky:1996ub,Kovchegov:1999yj}, that describe the small-$x$
evolution of hadronic wave functions. However, rather than in terms of
partonic densities, high-energy QCD evolution is more naturally
formulated in terms of correlators of Wilson lines as effective
degrees of freedom. The JIMWLK equation is equivalent to an infinite
set of coupled, non-linear evolution equations for all correlators of
the Wilson lines --$\,$also known as Balitsky's hierarchy. In the
limit of large number of colors ($N_c$) the hierarchy reduces to a
single equation --$\,$the BK equation$\,$-- for the correlator of two
Wilson lines or, equivalently, for the (imaginary part of the) dipole
scattering amplitude ${\cal N}$. As we shall explain in detail in
Section \ref{setup}, in the dipole model the small-$x$ dependence of
the different DIS cross sections is completely encoded in the dipole
scattering amplitude, and thus describable by the JIMWLK-BK equations.

Even though the JIMWLK equation comprises a richer physical input than
the BK equation,
%%%% Nestor
the latter
%it is this one which
%%%%
%%%% Gui 
% I prefer the original latter (it was misspelt though!)
%%%%
has become the most widely used tool to
study the small-$x$ dynamics. This is in part due to the relative
simplicity of the BK equation with respect to JIMWLK, whose solution demands the use of rather complicated numerical methods
\cite{Rummukainen:2003ns}. Further, the difference between the
solutions of the BK and JIMWLK equations turns out to be significantly
smaller, of order $0.1\%$ \cite{Rummukainen:2003ns}, than the a priori
expected ${\cal O}(1/N_c^2)$ corrections. The origin of the smallness
of the subleading-$N_c$ corrections have been investigated recently in
\cite{Kovchegov:2008mk}. For these reasons, here we will consider the
BK equation, rather than JIMWLK, as the starting point to analyse the
experimental data on the proton structure functions at small-$x$.

One of the first and most successful phenomenological applications of
saturation based ideas to the description of small-$x$ DIS data is due
to Golec-Biernat and W\"usthoff (GBW)
\cite{Golec-Biernat:1998js}. Their pioneering work relies on the use
of the dipole model in QCD \cite{Nikolaev:1990ja,Mueller:1989st}, 
together with a relatively simple model for the dipole-proton
scattering amplitude encoding the basic features of saturation, to
calculate the DIS total and diffractive lepton-proton cross
sections. In particular, the proton saturation scale was parametrized
as $Q_s^2(x)= (x_0/x)^{\lambda}$ GeV$^2$. Fits to HERA data yielded
$x_0=3\cdot 10^{-4}$ and $\lambda=0.288$. Several improvements of the
GBW model for the dipole scattering amplitude were proposed later on
in \cite{Iancu:2003ge,Kowalski:2003hm,Kowalski:2006hc,
  Albacete:2005ef,Goncalves:2006yt,Bartels:2002cj}. Very succinctly,
some of these works \cite{Iancu:2003ge} incorporated features of BFKL
dynamics and explicit impact parameter dependence in the scattering
amplitude \cite{Gotsman:2002yy,Kowalski:2003hm,Kowalski:2006hc},
whereas \cite{Bartels:2002cj} focused in including DGLAP evolution
into the model, which resulted in a improved fit to the higher $Q^2$
data. A first attempt of combining BK and DGLAP dynamics in the
description of DIS data was made in \cite{Gotsman:2002yy}. Finally,
the relation to heavy ion collisions was explored in
\cite{Albacete:2005ef,Goncalves:2006yt}. Overall, these works reported
an evolution speed compatible with the one obtained in the GBW model,
$\lambda\sim 0.2\div 0.3$.

A natural question arises of why the BK-JIMWLK equations, the most
solid theoretical tool available to describe the small-$x$ dynamics of
the dipole scattering amplitude and, in particular, the $x$-dependence
of the saturation scale, have not been directly applied to the study
of DIS small-$x$ data. The answer to this question is given by the
analytical \cite{Iancu:2002tr,Mueller:2002zm} and numerical
\cite{Armesto:2001fa,Braun:2003un,Albacete:2004gw} studies of the
leading-order (LO) BK equation. In these works the growth of the
saturation scale yielded by the LO BK equation was determined to be
$Q_s^2\sim x^{-\lambda_{LO}}$, with $\lambda_{LO}\simeq
4.88\,N_c\,\alpha_s/\pi$. Thus, the LO result predicts a much
faster growth of the saturation scale (and hence of DIS structure
functions) with decreasing $x$ than the one extracted
phenomenologically. This insufficiency of LO BK can only be
circumvented by introducing an unreasonably small value for the fixed
coupling, 
%%%% Gui
%making hopeless any attempt to describe experimental data.
rendering any attempt to describe experimental data far from meaningful.
%%%%

It has been a long-standing expectation that higher order corrections
to the original LO BK-JIMWLK equations could bring the theoretical
predictions closer to experimental observations. Indeed, numerical
estimates for the running coupling \cite{Braun:2003un,Albacete:2004gw}
and energy conservation corrections
\cite{Chachamis:2004ab,Albacete:2004gw} --$\,$both subleading physical
contributions to the LO kernel$\,$-- based on heuristic modifications
of the LO kernel indicated a significant reduction of the evolution
speed, thus pointing in the right direction. Moreover, running
coupling effects appeared to dominate the contribution to the
evolution kernel with respect to energy conservation effects
\cite{Albacete:2004gw}. However, it was not until recently that an
explicit first principle calculation of the running coupling
corrections to the evolution kernel was performed in
\cite{Balitsky:2006wa, Kovchegov:2006vj,Gardi:2006rp} by including $\as \, N_f$ corrections ($N_f$ being the number of flavors)
into the evolution kernel to all orders and by then completing $N_f$
to the one-loop QCD beta-function.
%The
%calculations in \cite{Balitsky:2006wa,Kovchegov:2006vj} proceeded by
%including $\as \, N_f$ corrections ($N_f$ being the number of flavors)
%into the evolution kernel to all orders and by then completing $N_f$
%to the one-loop QCD beta-function via replacing $N_f \rightarrow - 6
%\pi \beta_2$, with $\beta_2=(11N_c-2N_f)/(12\pi)$. The calculation of
%the $\as \, N_f$ corrections is particularly simple in the $s$-channel
%light-cone perturbation theory (LCPT) formalism used to derive the BK
%and JIMWLK equations: there $\as \, N_f$ corrections are solely due to
%chains of quark bubbles placed onto the $s$-channel gluon lines. The
%calculation in \cite{Gardi:2006rp} relied instead on the use of
%dispersive methods, arriving at the same results as in the
%perturbative calculation in \cite{Kovchegov:2006vj}.
%
The numerical study of the BK equation at all orders in $\alpha_s
N_f$, performed in \cite{Albacete:2007yr}, reported a significant
slowdown of the evolution speed with respect to the solutions of the
LO equation, hence rising the hopes that the improved equation might
become a useful phenomenological tool. In its first successful
application it was used to describe the energy and rapidity
dependences of particle multiplicities produced in nucleus-nucleus
collisions at the Relativistic Heavy Ion Collider (RHIC) at the BNL
\cite{Albacete:2007sm}.

%%%% Gui
% This is explained in sect.2. in my opinion it could be removed from here
%It is important to note that the mentioned $\alpha_s\, N_f$
%corrections contain not only running coupling corrections to the
%kernel, but also contributions from conformal, non-running coupling
%terms originating from new physical channels, namely quark-antiquark
%final states. As discussed in great detail in \cite{Albacete:2007yr},
%the separation between those two contributions is ambiguous and, not
%surprisingly, it was performed differently in \cite{Balitsky:2006wa}
%and \cite{Kovchegov:2006vj}. Thus, the kernel including running
%coupling corrections is subject to scheme dependence on the choice of the separation between running and subtraction contributions.
%%%%
Significant progress has also been made recently in the determination
of subleading physical effects, other than running coupling
corrections, to the LO BK equation, namely the inclusion of pomeron
loops (see e.g.  \cite{Albacete:2006uv,Altinoluk:2009je} and
references therein), finite-$N_c$ corrections \cite{Kovchegov:2008mk}
or the determination of the complete next-to-leading evolution kernel
\cite{Balitsky:2008zz} to the BK equation. However, our current
understanding indicates that the running coupling effects are dominant
with respect to pomeron loops (or particle number fluctuations)
\cite{Dumitru:2007ew}
%%%% Gui
% \cite{Abreu:2007kv} 
% is this reference right???????????, the only such statement i know is from edmond et al, 0706:2540
% i replaced it by adding the above
%%%%
or finite $N_c$ corrections
\cite{Kovchegov:2008mk}. We will therefore limit ourselves in the
present work to the analysis of DIS small-$x$ data via the BK equation
including {\it only} running coupling corrections.

The first goal of this paper is to prove the ability of the BK
equation including running coupling corrections to account for the
small-$x$ behavior of the total, $F_2$, and longitudinal, $F_L$,
structure functions measured in DIS experiments (a first step in this
direction, yet unpublished, was reported in \cite{WT}). To that end we
shall devise a global fit to the available experimental data with
$x\leq x_0=10^{-2}$ and for all values of $Q^2$. Analogously to
previous works, our starting point will be the dipole model of
QCD. The main novelty of our work is that the dipole-proton scattering
amplitude, instead of being modeled, is calculated via numerical
solutions of the BK equation including running coupling
corrections. The free parameters in our fit, to be detailed in Section
\ref{setup}, are those related to the parametrization of the initial
condition for the evolution, a global coefficient that sets the
normalization and a constant which relates the running of the coupling
in momentum space to that in dipole size. As we show in Section
\ref{f2fl}, the fits yield a good $\chi^2/{\rm d.o.f.}\le 1.1$,
thus demonstrating that such partial improvement of the LO BK equation
suffices to reconcile the theoretical predictions with experimental
results. Further, in Section \ref{predictions} with all the free
parameters fixed by the global fit of available data, we make
predictions for the same observables at much smaller values of
$x$. Such predictions are completely driven by non-linear QCD dynamics
and could be directly tested at the proposed Electron-Ion Collider
(EIC) \cite{eic} or Large Hadron-electron Collider (LHeC) \cite{lhec2}
experimental facilities, where values of $x$ as low as $x\sim 10^{-7}$
for $Q^2\sim 1$ GeV$^2$ could be reached.

Second, the upcoming LHC experimental programs in proton-proton,
proton-nucleus and nucleus-nucleus demand a detailed knowledge of
hadronic wave functions or parton density functions (PDF) at very
small $x$ as an input for the calculation of many different
observables (see, for instance, the discussions in
\cite{Abreu:2007kv,Dittmar:2009ii}).  While global PDF fits provide a
description of currently available data, additional theoretical input
is needed in order to safely extrapolate towards values of $x$ so far
unexplored empirically
%%%% Gui
and for which additional saturation effects appear unavoidable.
%%%%  
A similar situation is found in cosmic rays
physics \cite{d'Enterria:2008jk,Armesto:2007tg}, where the highest
center-of-mass energies reached in primary collisions are simply
unattainable in accelerator experiments in the foreseeable future. In
this work we set the ground for a systematic program oriented to
provide parameter-free extrapolations of the dipole amplitudes (both
for proton and nuclei) to very small values of $x$ based on first
principle calculations. Parametrizations of the dipole-proton
scattering amplitudes down to very small $x$ based on the results of
this work 
%%%% Gui
are publicly available 
%shall be made public 
%%%%
through simple numeric routines
\cite{pweb}.

%%%%%%%%%%%%%%%%%%%%%%%%%%%%%%%%%%%%%%%%%%%%%%%%%%%%%%%%%%%%%%%%%%%%%%%%%%%%%

\section{Setup} \label{setup}
In this section we briefly review, 
%%%% Gui
in a self contained manner,
%%%%
the main ingredients needed for the
calculation of the inclusive and longitudinal DIS structure functions.
\subsection{Dipole model}
\label{dm}

At $x\ll 1$, the inclusive structure function of DIS can be expressed
as
\begin{equation}
F_2(x,Q^2)=\frac{Q^2}{4\,\pi^2\alpha_{em}}\left(\sigma_T+\sigma_L\right)\,,
\label{f2}
\end{equation}
where $\alpha_{em}$ is the electromagnetic coupling and $\sigma_{T,L}$
stands for the virtual photon-proton cross section for transverse ($T$)
and longitudinal ($L$) polarizations of the virtual  photon. The longitudinal
structure function is obtained by considering only the longitudinal
contribution:
\begin{equation}
F_L(x,Q^2)=\frac{Q^2}{4\,\pi^2\alpha_{em}}\,\sigma_L\,.
\label{fl}
\end{equation}
It is well known that at high energies or small $x$ (where the
coherence length of the virtual photon fluctuation $l_c\approx (2m_N
x)^{-1}\simeq 0.1/x \ {\rm fm} \gg R_N$, with $m_N$ and $R_N$ the
proton mass and radius respectively), and using light-cone
perturbation theory, the total virtual photon-proton cross section can
be written as the convolution of the light-cone wave function squared
for a virtual photon to fluctuate into a quark-antiquark dipole,
$|\Psi_{T,L}|^2$, and the imaginary part of the dipole-target
scattering amplitude, ${\cal N}$. For transverse and longitudinal
polarizations of the virtual photon one writes
\cite{Nikolaev:1990ja,Mueller:1989st}:
\begin{equation}
  \sigma_{T,L}(x,Q^2)=\int_0^1 dz\int d{\bf b} \,d{\bf r}\,\vert
  \Psi_{T,L}(z,Q^2,{\bf r})\vert^2\,
  {\cal N}({\bf b},{\bf r},x)\,,
\label{dm1}
\end{equation}
where $z$ is the fraction of longitudinal momentum of the photon
carried by the quark, ${\bf r}$ is the transverse separation between
the quark and the antiquark and ${\bf b}$ the impact parameter of the
dipole-target collision (henceforth boldface notation indicates
two-dimensional vectors). The wave functions $\vert \Psi_{T,L}\vert^2$
for the splitting of the photon into a $q\bar{q}$ dipole are
perturbatively computable within QED. We refer the reader to
e.g. \cite{Golec-Biernat:1998js} for explicit expressions to lowest
order in $\alpha_{em}$. All the information about the strong
interactions --$\,$along with all $x$-dependence$\,$-- in \eq{dm1} is
encoded in the dipole-proton scattering amplitude, ${\cal N}({\bf
  b},{\bf r},x)$. Although this quantity is a genuinely
non-perturbative object, its evolution towards smaller values of $x$
can be studied perturbatively via the BK equation. On the contrary,
its impact parameter dependence cannot be studied by means of the
perturbative BK equation, since it is governed by long distance,
non-perturbative physics. To circumvent this theoretical limitation we
will resort to the translational invariance approximation (also used
in \cite{Golec-Biernat:1998js}), which regards the proton as
homogeneous in the transverse plane. Under this approximation the
virtual photon-proton cross section \eq{dm1} can be rewritten as
follows:
\begin{equation}
  \sigma_{T,L}(x,Q^2)=\sigma_0\,\int_0^1 dz\int \,d{\bf r}\,\vert
  \Psi_{T,L}(z,Q^2,{\bf r})\vert^2\,
  {\cal N}(r,Y)\,,
\label{dm2}
\end{equation}
where $r=\vert {\bf r} \vert$ is the dipole size (the notation $v
\equiv |\bf{v}|$ for all the 2-dimensional vectors will be also
employed throughout the rest of the paper) and $\sigma_0$ is a
dimensionful constant resulting from the ${\bf b}$ integration that
sets the normalization --$\,$this will be one of the free parameters
in our fits.
Note that this result relies on the assumption that a factorized structure of $x,r$ and $b$ dependences remains unchanged throughout the evolution. In this case the parameter $\sigma_0$ is related to the $t$-dependence in diffractive events, see e.g. \cite{Marquet:2007nf}. On the other hand, this factorized structure may be assumed solely for the initial condition, while small-$x$ evolution is performed, in the translational-invariant approximation, separately for every impact parameter (as done e.g. for nuclei in \cite{Levin:2001et,Armesto:2001vm}). This results in a $\sigma_0$ varying (increasing) with energy \cite{Kopeliovich:1999am}. We leave this latter aspect for future studies.

\subsection{BK equation with running coupling}
\label{bk}
The primary physical mechanism driving the small-$x$ evolution of the
dipole scattering amplitude is the emission of soft gluons off either
the quark or the antiquark in the original dipole. The leading order
BK equation resumming the corresponding $\alpha_s\,\ln(1/x)$
contributions to all orders reads
\begin{eqnarray}
  \frac{\partial{\cal{N}}(r,Y)}{\partial\,Y}&=& \int d{\bf r_1}\,
  K^{{\rm LO}}({\bf r},{\bf r_1},{\bf r_2})\nonumber \\
 &\times&
 \left[{\cal N}(r_1,Y)+{\cal N}(r_2,Y)-{\cal N}(r,Y)-
    {\cal N}(r_1,Y)\,{\cal N}(r_2,Y)\right]\,,
\label{bklo}
\end{eqnarray}
with the evolution kernel given by
\begin{equation}
  K^{{\rm LO}}({\bf r},{\bf r_1},{\bf r_2})=\frac{N_c\,\alpha_s}{2\pi^2}
  \,\frac{r^2}{r_1^2\,r_2^2}\,,
\label{kballo}
\end{equation}
%%%% Gui
% removed scale from \alpha_s, we had already spotted this misprint
%%%%
and ${\bf r_2}={\bf r}-{\bf r_1}$. Here, $Y\!=\!\ln (x_0/x)$ is the
rapidity variable and $x_0$ is the value of $x$ where the evolution
starts, which should be small enough for the dipole model to be
applicable. In our case $x_0=0.01$ will be the highest experimental
value of $x$ included in the fit.

The
calculations in \cite{Balitsky:2006wa,Kovchegov:2006vj} proceeded by
including $\as \, N_f$ corrections ($N_f$ being the number of flavors)
into the evolution kernel to all orders and by then completing $N_f$
to the one-loop QCD beta-function via replacing $N_f \rightarrow - 6
\pi \beta_2$, with $\beta_2=(11N_c-2N_f)/(12\pi)$.
The calculation of
the $\as \, N_f$ corrections is particularly simple in the $s$-channel
light-cone perturbation theory (LCPT) formalism used to derive the BK
and JIMWLK equations: there $\as \, N_f$ corrections are solely due to
chains of quark bubbles placed onto the $s$-channel gluon lines, as sketched in Fig. \ref{nlodip}A.
Importantly, at the
same degree of accuracy a new physical channel is opened, namely the
emission of a quark-antiquark pair, instead of a gluon, as depicted in
Fig. \ref{nlodip}B.
The
calculation in \cite{Gardi:2006rp} relied instead on the use of
dispersive methods, arriving at the same results as in the
perturbative calculation in \cite{Kovchegov:2006vj}.
%
%In short, the inclusion of $\alpha_s\,N_f$ corrections consists in
%dressing the propagator of such newly emitted gluon with quark loops
%to all orders, as sketched in Fig. \ref{nlodip}A. Importantly, at the
%same degree of accuracy a new physical channel is opened, namely the
%emission of a quark-antiquark pair, instead of a gluon, as depicted in
%Fig. \ref{nlodip}B.

Neglecting the impact parameter dependence, the improved BK evolution
equation for the dipole scattering amplitude obtained after resumming
the subleading $\alpha_s N_f$ corrections to all orders in
\cite{Balitsky:2006wa,Kovchegov:2006vj} can be written in the
following, rather general form \cite{Albacete:2007yr}:
\begin{equation}
  \frac{\partial {\cal N}(r,Y)}{\partial Y}=
  {\cal R}[{\cal N}]-{\cal S}[{\cal N}]\,,
\label{bkfun} 
\end{equation}
where both ${\cal R}$ and ${\cal S}$ are functionals of the dipole
scattering amplitude, ${\cal N}$. The first, {\it running coupling},
term ${\cal R[{\cal N}]}$ in \eq{bkfun} gathers all the $\alpha_s \,
N_f$ factors needed to complete the QCD beta function, resulting in a
functional form identical to the LO one but involving a modified
kernel which provides the scale setting for the running of the
coupling. In turn, the second term in the r.h.s. of \eq{bkfun}, ${\cal
  S}[{\cal N}]$, the {\it subtraction} term, accounts for conformal,
non running-coupling contributions.
%%%%%%%%%%%%%%%%%%%%%%%%%%%%%
\FIGURE{
%\begin{figure}[ht]
%\begin{center}
\includegraphics[height=3.7cm]{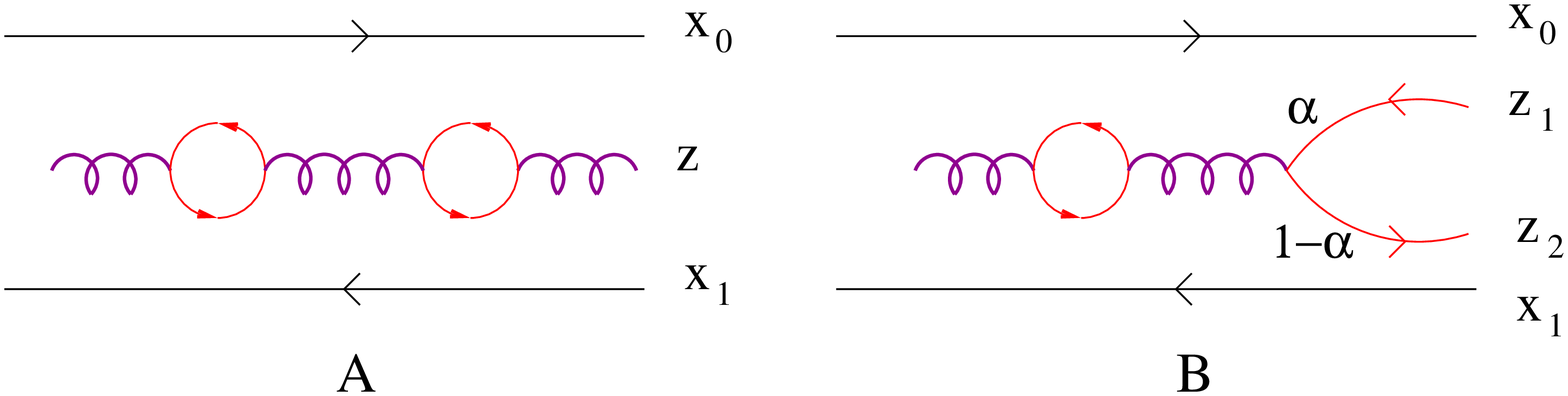}%,width=14.5cm]{c1.eps}
%\end{center}
\caption{Schematic representation of the diagrams contributing to the
  evolution to all orders in $\alpha_sN_f$. The s-channel gluon line
  can be attached to either the quark (upper line) or the antiquark
  (lower line).}
\label{nlodip}
%\end{figure}
}
%%%%%%%%%%%%%%%%%%%%%%%%%%%%%

It would be erroneous to identify the gluon and quark-antiquark
emission channels with the running and subtraction terms in \eq{bkfun}
respectively. Importantly, the quark-antiquark channel contains a
logarithmic ultra-violet (UV) divergence associated to the emission of
a zero size pair which, in the large-$N_c$ limit, is indistinguishable
from one gluon emission and therefore contributes to the running of
the coupling on an equal footing. The emission of finite size
quark-antiquark pairs is UV finite and does not contribute to the
running of the coupling. Thus, the decomposition of the evolution
kernel into {\it running} and {\it subtraction} contributions,
although constrained by unitarity arguments, is not unique. This is
due to the fact that there is some freedom in the way in which the UV
divergence can be singled out from the {\it conformal} one, so in
order to perform a decomposition like the one in \eq{bkfun} a precise
separation scheme needs to be specified. Not surprisingly, the
separation schemes proposed in \cite{Balitsky:2006wa} and
\cite{Kovchegov:2006vj} were different. For a detailed discussion on
this subject we refer the reader to \cite{Albacete:2007yr}.

In this work we will consider only the {\it running} term in the
evolution kernel. Ideally one would like to include the {\it
  subtraction} piece of the evolution kernel in practical applications
as this would eliminate the uncertainty associated to the scheme
choice and would provide a richer physical description of the
small-$x$ evolution of the dipole scattering amplitude. Unfortunately,
its numerical evaluation \cite{Albacete:2007yr} demands a very large
computing time. For a global fit like the one presented in this work,
in which the evolution is performed $\sim 10^3$ times, such computing
time is simply unaffordable. On the other hand, as shown in
\cite{Albacete:2007yr} the contribution to the complete evolution
kernel stemming from the {\it subtraction} term is systematically
smaller --$\,$and negligible at high rapidities$\,$-- than the one
arising from the {\it running} term. In particular, we will follow the
prescription proposed by Balitsky in \cite{Balitsky:2006wa} to single
out the {\it running} term since, as demonstrated in
\cite{Albacete:2007yr}, such choice minimizes the contribution to the
evolution of the {\it subtraction} term, neglected in what follows,
with respect to the separation scheme proposed in
\cite{Kovchegov:2006vj}.

Finally, after dropping the subtraction term from \eq{bkfun}, the BK
evolution equation including {\it only} running coupling corrections
reads
\begin{equation}
  \frac{\partial {\cal N}(r,Y)}{\partial Y}=
  {\cal R}^{{\rm Bal}}[{\cal N}]\,, 
\label{bk1}
\end{equation}
where the running coupling functional is identical to the LO equation:
\begin{eqnarray}
  {\cal R}^{{\rm Bal}}[{\cal N}]&=& \int d{\bf r_1}\,
  K^{{\rm Bal}}({\bf r},{\bf r_1},{\bf r_2})\nonumber\\ 
&\times&  \left[{\cal N}(r_1,Y)+{\cal N}(r_2,Y)-{\cal N}(r,Y)-
    {\cal N}(r_1,Y)\,{\cal N}(r_2,Y)\right]\,,
\label{bk2}
\end{eqnarray}
but with a modified evolution
kernel that includes running coupling corrections.  Using Balitsky's
prescription, the kernel for the {\it running} term reads
\cite{Balitsky:2006wa}
\begin{equation}
  K^{{\rm Bal}}({\bf r},{\bf r_1},{\bf r_2})=\frac{N_c\,\alpha_s(r^2)}{2\pi^2}
  \left[\frac{r^2}{r_1^2\,r_2^2}+
    \frac{1}{r_1^2}\left(\frac{\alpha_s(r_1^2)}{\alpha_s(r_2^2)}-1\right)+
    \frac{1}{r_2^2}\left(\frac{\alpha_s(r_2^2)}{\alpha_s(r_1^2)}-1\right)
  \right]\,.
\label{kbal}
\end{equation}

%%%%%%%%%%%%%%%%%%%%%%%%%%%%%%%%%%%%%%%%%%%%%%%%%%%%%%%%%%%%%%%%%%%%%%%%%%%%%%%
\subsection{Regularization of the infrared dynamics}
\label{ir}
The BK equation is an integro-differential equation that involves
integration over all available phase-space for soft gluon emission. In
the running coupling case, Eqs. (\ref{bk1}-\ref{kbal}), the coupling
has to be evaluated at arbitrarily large values of the dipole size
(small gluon momentum), and a regularization prescription to avoid the
Landau pole becomes necessary. A celebrated feature of the BK equation
is its ability to fix \cite{Golec-Biernat:2001if} the problem of
infra-red diffusion characteristic of its linear counterpart, the BFKL
equation. The non-linear terms in the BK equation ensure that the
dynamics in the phase space region within the unitarity limit,
i.e. for $r\gg1/Q_s$, is frozen. Such feature is shared by both the LO
and running coupling BK equations, since it is ultimately rooted in
the non-linear combination of ${\cal N}$'s in the r.h.s. of \eq{bk2},
which is identical in both cases. Thus, if $Q_s$ is perturbatively
large, $Q_s\gg\Lambda_{QCD}$, then all the relevant dynamics takes
place deep in the ultra-violet region of the phase space, $r\leq
1/Q_s$. In such scenario the details about the regularization of the
running coupling in the infra-red become irrelevant for the result of
the evolution.

Unfortunately, we can anticipate that such will not be the case in
this work. Taking the results by Golec-Biernat and W\"usthoff
\cite{Golec-Biernat:1998js} as a guidance, one can estimate that the
proton saturation scale at the largest values of Bjorken-$x$ to be
considered in this work, $x\sim 10^{-2}$, is of the order of
$Q_s^2(x\!=\!10^{-2})\approx (3\cdot 10^{-4}/10^{-2})^{0.288}$
GeV$^2\simeq 0.36$ GeV$^2$. The fits to be presented in Section
\ref{results} yield even smaller values of the initial saturation
scale of the proton.  Although larger than $\Lambda_{QCD}^2$, such
values for the initial scale are not large enough to avoid sensitivity
to the infra-red (IR) dynamics. Actually, the detailed study of the
infrared-renormalon ambiguities carried out in \cite{Gardi:2006rp}
demonstrated that the sensitivity of the solutions of the evolution
equation to several different prescriptions used to regularize the
coupling is relatively large even for initial saturation scales as
large as $Q_s^2\sim1\div2$ GeV$^2$.  On the bright side, theoretical
studies of the Schwinger-Dyson equations for the gluon propagator in
the IR and lattice QCD results (see e.g.
\cite{Cornwall:1981zr,Aguilar:2008xm} and references therein) indicate
that the strong coupling freezes to a constant value, $\alpha_{fr}$,
in the IR. Moreover, the value at which the coupling freezes has been
determined to be $\alpha_{fr}\sim 0.5\div0.7$. While these results are
somewhat controversial and yet subject to discussion in the
literature, in particular the very definition of an infrared coupling,
we will take them as a guidance to regularize the IR
dynamics. Otherwise, our prescription can be regarded as purely
phenomenological.

Thus, for small dipole sizes $r< r_{fr}$, with
$\alpha_s(r_{fr}^2)\equiv\alpha_{fr}=0.7$, we shall evaluate the
running coupling according to the usual one-loop QCD expression:
\begin{equation}
\alpha_s(r^2)=\frac{12\pi}{\left(11N_c-2N_f\right)\,\ln\left
    (\frac{4\,C^2}{r^2 \Lambda_{QCD}^2}\right)}\,,
\label{alpha}
\end{equation}
with $N_f=3$, whereas for larger sizes, $r>r_{fr}$, we freeze the
coupling to the fixed value $\alpha_{fr}=0.7$. We take
$\Lambda_{QCD}=0.241$ GeV, such that $\alpha_s(M_{Z})=0.1176$, with
$M_{Z}$ the mass of the $Z$ boson. The factor $C^2$ under the
logarithm in \eq{alpha} will be one of the free parameters in the
fit. It reflects the uncertainty in the Fourier transform from
momentum space, where the original calculation of $\alpha_sN_f$
corrections was performed, to coordinate space. Alternatively, we
could have fixed $C^2$ to the value suggested in
\cite{Kovchegov:2006vj}, $e^{-5/3-2\gamma_E}$, and chosen either
$\Lambda_{QCD}$ or $\alpha_{fr}$ as the free parameters controlling
the IR dynamics. Indeed, we have checked that such choices yield
equally good fits as those presented in Section \ref{results} without
changing much the value of the other free parameters. However, both
$\alpha_{fr}$ and, specially, $\Lambda_{QCD}$, are more tightly
constrained from both theoretical and phenomenological studies than
$C^2$.

\subsection{Initial conditions for the evolution}
\label{ic}

Finally we have to specify the initial condition (i.c.) for the
evolution or, equivalently, the precise shape of the proton
unintegrated gluon distribution (UGD), $\phi(x,k)$, at the
highest experimental value of Bjorken-$x$ included in the fit, $x_0=0.01$
(which, by definition, corresponds to rapidity $Y\!=\!0$). The
UGD is related to the dipole scattering amplitude via a Fourier
transform:
\begin{equation} 
  \phi(x,k)=\int \frac{d{\bf r}}{2\,\pi \,r^2}\,e^{i\,{\bf
      k}\cdot {\bf r}}\,{\cal N}(x,r)\,. 
\end{equation}
This is a genuinely non-perturbative object which needs to be
modeled. We will consider two different families of initial
conditions. The first one is inspired in the original GBW ansatz
\cite{Golec-Biernat:1998js} for the dipole scattering amplitude and
parametrized in the following way: \beq {\cal N}^{GBW}(r,Y\!=\!0)=
1-\exp{\left[-\left(\frac{r^2\,Q_{s\,0}^2}{4}\right)^{\gamma\,}\right]}\,.
\label{gbw}
\eeq The second family of initial conditions \cite{McLerran:1997fk}
follows closely the McLerran-Venugopalan (MV) model: \beq {\cal
  N}^{MV}(r,Y\!=\!0)=1-\exp{\left[-\left(\frac{r^2Q_{s\,0}^{2}}{4}\right)^{\gamma}
    \ln{\left(\frac{1}{r\,\Lambda_{QCD}}+e\right)}\right]}\, ,
\label{mv}
\eeq where $Q_{s\,0}^2$ is the initial saturation scale squared in
both cases.

Eqs. (\ref{gbw}) and (\ref{mv}) differ with respect to the
original GBW and MV models in the inclusion of an anomalous dimension,
$\gamma$, which will be another of the free parameters in the fit. The
GBW and MV functional forms are recovered by setting $\gamma\!=\!1$ in
\eq{gbw} and \eq{mv} respectively. The anomalous dimension controls
the slope of the scattering amplitude in the transition from the
dilute region to the black disk region. The main difference between MV
and GBW i.c. is their different UV behaviour, which is more easily
appreciated in momentum space. For $\gamma=1$ and large transverse
momenta $k$, the UGD resulting from the MV i.c. falls off as
$\phi^{MV}\sim 1/k^2$, as expected from rather general perturbative
considerations, while the GBW i.c. falls off exponentially,
$\phi^{GBW}\sim \exp{(-k^2/Q_s^2)}$.  It is well known that the
solutions of the BK equation, both at LO and including higher order
corrections, do not respect the relatively simple functional forms in
\eq{gbw} and \eq{mv}. On the contrary, they can be roughly
characterized by an $r$- and $Y$-dependent anomalous dimension,
$\gamma(r,Y)$, with $\gamma\to 1$ for $r\to 0$. Clearly a constant
value of $\gamma\ne1$ would not respect such condition. However, the
main contribution to the DIS cross section given by \eq{dm2}
originates from the region $1/Q\lesssim r\lesssim 1/Q_s$.
%, where ${\cal N}\sim1$.
The contribution from the dilute UV region $r<1/Q$ is much smaller and
therefore we will not consider additional refinements of the initial
conditions in \eq{gbw} and \eq{mv}, which would come at the prize of
adding new, spurious parameters into the fit.  [Actually, the results
of the fit shows that for the GBW i.c. the preferred value is
$\gamma=1$, so it will be fixed for this initial condition.]  Finally,
the constant term under the logarithm in the MV initial condition,
$e$, has been added to regularize the exponent for large values of
$r$.

\subsection{Summary of the theoretical setup and free parameters}
In summary, we will calculate the total DIS inclusive and longitudinal
structure functions according to the dipole model under the
translational invariant approximation \eq{dm2}.  The small-$x$
dependence is completely described by means of the BK equation
including running coupling corrections, Eqs. (\ref{bk1}-\ref{kbal}),
for which two different initial conditions GBW and MV,
Eqs. (\ref{gbw}) and (\ref{mv}), are considered. All in all, the free
parameters to be fitted to experimental data are:
\begin{itemize}
\item $\sigma_0\,$: The total normalization of the cross section in \eq{dm2}.
\item $Q_{s\,0}^2\,$: The saturation scale of the proton at the highest
  experimental value of Bjorken-$x$ included in the fit, $x_0=10^{-2}$, in Eqs. (\ref{gbw}) and (\ref{mv}).
\item $C^2$: The parameter relating the running of the
  coupling in momentum space to the one in dipole size in \eq{alpha}.
\item $\gamma\,$: The anomalous dimension of the initial condition for
  the evolution in Eqs. (\ref{gbw}) and (\ref{mv}). As discussed in
  Section \ref{results}, in some cases (GBW) $\gamma$ can be fixed to
  1, obtaining equally good fits to data than when it is considered
  a free parameter.
\end{itemize}

\section{Numerical method and experimental data}
\label{numer}
The experimental data included in the fit to $F_2(x,Q^2)$ have been
collected by the E665 \cite{Adams:1996gu} (FNAL), the NMC
\cite{Arneodo:1996qe} (CERN-SPS), the H1
\cite{Abt:1993cb,Ahmed:1995fd,
  Aid:1996au,Adloff:1997mf,Adloff:1999ah,Adloff:2000qk} (HERA) and the
ZEUS \cite{Derrick:1993fta,
  Derrick:1994sz,Derrick:1995ef,Derrick:1996hn,Breitweg:1997hz,Breitweg:1998dz,
  Breitweg:2000yn,Chekanov:2001qu} (HERA) experimental
Collaborations. We have considered data for $x\leq10^{-2}$ and for all
available values of $Q^2$, $0.045 \ {\rm GeV}^2\le Q^2 \le 800 \ {\rm
  GeV}^2$.

The only published direct measurements of the longitudinal structure
function $F_L(x,Q^2)$ were obtained recently by the H1
\cite{:2008tx} and ZEUS \cite{Collaboration:2009na} Collaborations, and they are {\it not} included in the fit.

All in all, 847 data points are included.
Statistical and systematic uncertainties were added in quadrature, and
normalization uncertainties not considered. [A more involved treatment
separating uncorrelated and correlated/normalization errors could be
done only at the expense of adding one more fitting parameter for each
of the 17 data sets used, thus making the minimization task impossible
due to CPU-time requirements.]  Since the minimization algorithms
require a large number of calls to the function we have implemented a
parallelization of the numeric code. Finally, the BK
evolution equation including running coupling corrections is solved
using a Runge-Kutta method of second order with rapidity step $\Delta
h_y=0.05$, see further details in \cite{Albacete:2007yr}.

In order to smoothly go to photoproduction, we follow
\cite{Golec-Biernat:1998js} and use the redefinition of the Bjorken
variable
\begin{equation}
\tilde{x}=x\,\left(1+\frac{4m_f^2}{Q^2}\right),
\label{redef}
\end{equation}
with $m_f=0.14$ GeV for the three light flavors we consider in \eq{dm2}.

\section{Results} \label{results} 
\subsection{Fits to $F_2$ and description of $F_L$}
\label{f2fl}
The values of the free parameters obtained from the fits to data for
the two different initial conditions, GBW and MV, are presented in
Table 1. A partial comparison between the experimental data
\cite{Adams:1996gu,Arneodo:1996qe,Abt:1993cb,Ahmed:1995fd,
  Aid:1996au,Adloff:1997mf,Adloff:1999ah,Adloff:2000qk,Derrick:1993fta,
  Derrick:1994sz,Derrick:1995ef,Derrick:1996hn,Breitweg:1997hz,Breitweg:1998dz,
  Breitweg:2000yn,Chekanov:2001qu} and the results of the fit for
$F_2(x,Q^2)$ is shown in Fig. \ref{figf2}.

\TABLE{
%\begin{table}[h]
%\begin{center}
\begin{tabular}{c|c|c|c|c|c}
  Initial condition & $\sigma_0$ (mb) & $Q_{s0}^2$ (GeV$^2$) &
  $C^2$ &$\gamma$ & $\chi^2/{\rm d.o.f.}$ \\
  \hline
  GBW & 31.59 & 0.24 & 5.3 & 1 (fixed) & 916.3/844=1.086 \\
  \hline
  MV & 32.77 & 0.15 & 6.5 & 1.13 & 906.0/843=1.075\\
   \hline
\end{tabular}
\label{restab}
\caption{Values of the fitting parameters from the fit to $F_2(x,Q^2)$
  data from \cite{Adams:1996gu,Arneodo:1996qe,Abt:1993cb,Ahmed:1995fd,
    Aid:1996au,Adloff:1997mf,Adloff:1999ah,Adloff:2000qk,Derrick:1993fta,
    Derrick:1994sz,Derrick:1995ef,Derrick:1996hn,Breitweg:1997hz,Breitweg:1998dz,
    Breitweg:2000yn,Chekanov:2001qu} with $x\leq10^{-2}$ and for all
  available values of $Q^2$, $0.045 \ {\rm GeV}^2\le Q^2 \le 800 \
  {\rm GeV}^2$.}
%\end{center}
%\end{table}
}

%%%%%%%%%%%%%%%%%%%%%%%%%%%%%
\FIGURE{
%\begin{figure}[htp]
%\begin{center}
\includegraphics[width=17cm]{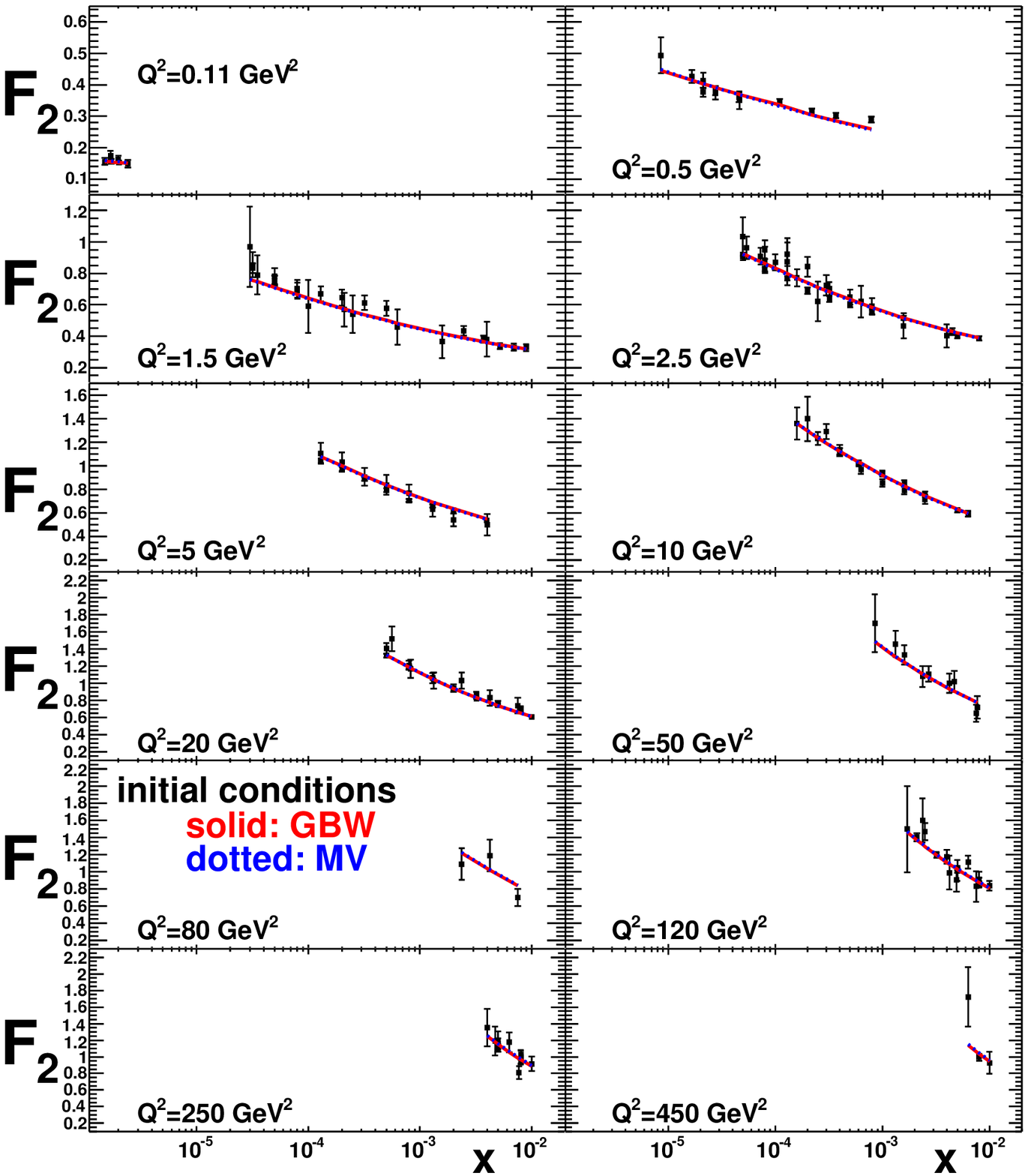}
%\end{center}
\caption{Comparison between a selection of experimental data
  \cite{Adams:1996gu,Arneodo:1996qe,Abt:1993cb,Ahmed:1995fd,
    Aid:1996au,Adloff:1997mf,Adloff:1999ah,Adloff:2000qk,Derrick:1993fta,
    Derrick:1994sz,Derrick:1995ef,Derrick:1996hn,Breitweg:1997hz,Breitweg:1998dz,
    Breitweg:2000yn,Chekanov:2001qu} and the results from the fit for
  $F_2(x,Q^2)$. Solid red lines correspond to GBW i.c., and dotted
  blue ones to MV i.c. The error bars correspond to statistical and
  systematic errors added in quadrature.}
\label{figf2}
%\end{figure}
}
%%%%%%%%%%%%%%%%%%%%%%%%%%%%%

On the other hand, $F_L(x,Q^2)$ offers an additional constrain on the
gluon distribution and is expected to have more discriminating power
on different approaches, particularly in the low-$Q^2$ region
\cite{Thorne:2008aj}. In Fig. \ref{figfl} we show a comparison between
experimental data \cite{:2008tx,Collaboration:2009na} and our predictions for $F_L(x,Q^2)$.

%%%%%%%%%%%%%%%%%%%%%%%%%%%%%
\FIGURE{
%\begin{figure}[ht]
%\begin{center}
\centerline{\includegraphics[width=12cm]{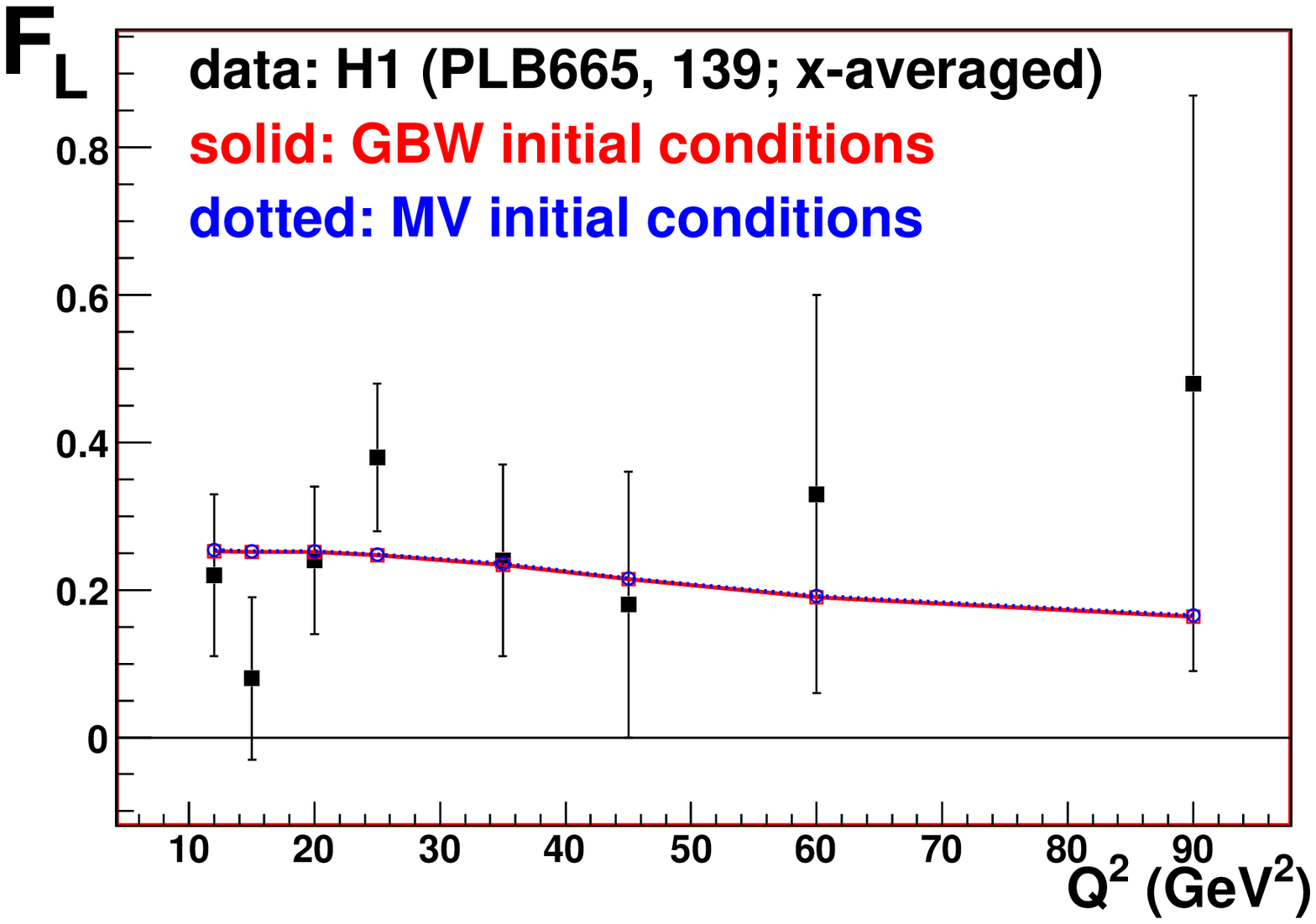}}
\centerline{\includegraphics[width=12cm]{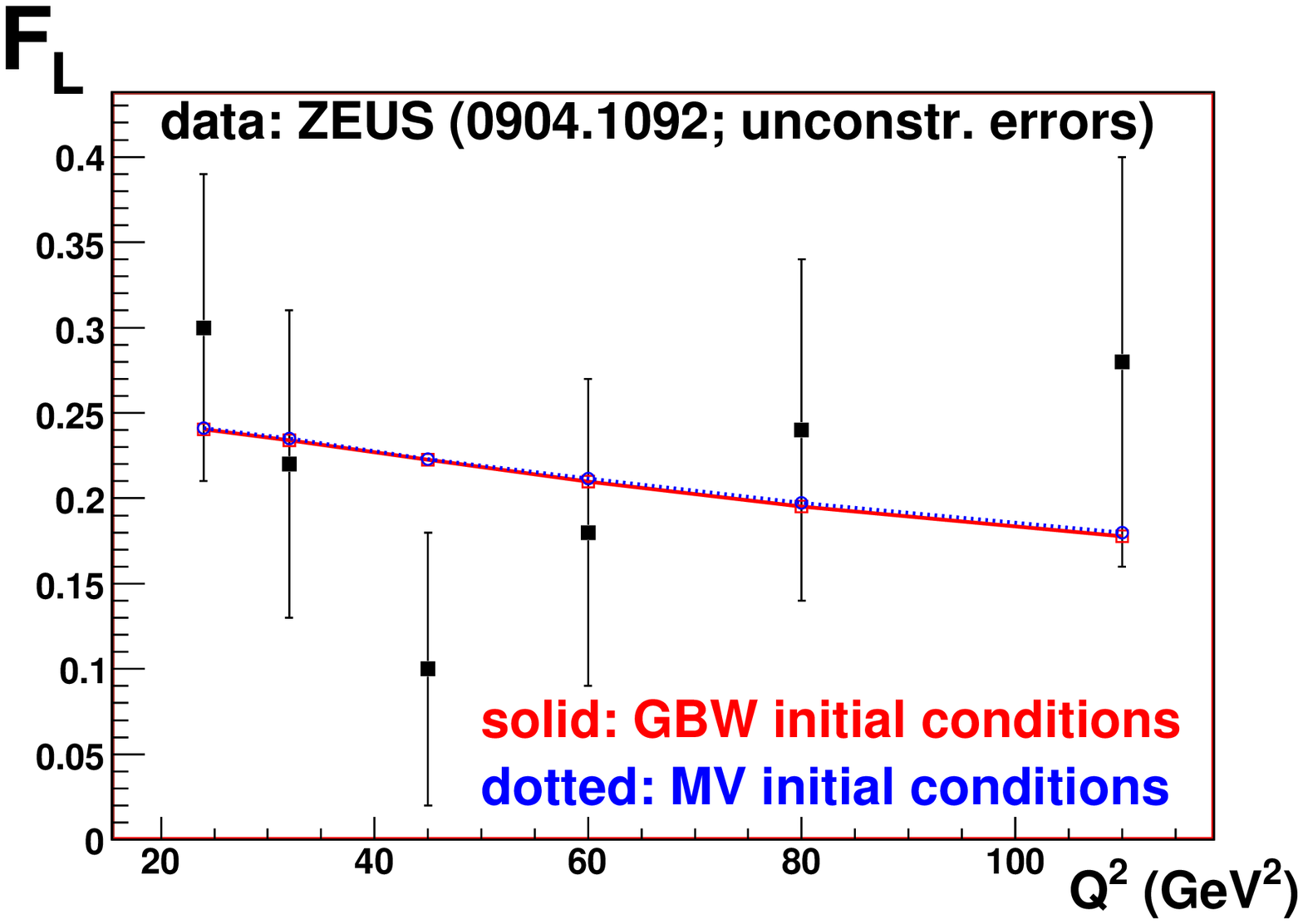}}
%\end{center}
\caption{Comparison between experimental data from the H1 \cite{:2008tx} (upper plot)  and ZEUS \cite{Collaboration:2009na} (lower plot) Collaborations and the
  predictions of our model for $F_L(x,Q^2)$. Red solid lines and open
  squares correspond to GBW i.c., and blue dotted lines and open
  circles to MV i.c. The theoretical results have been computed at the
  same $\langle x\rangle$ as the experimental data, and then joined by
  straight lines. The error bars correspond to statistical and
  systematic errors added in quadrature for those data coming from \cite{:2008tx}, while they correspond to the error quoted for the unconstrained fit for those data coming from \cite{Collaboration:2009na}.}
\label{figfl}
%\end{figure}
}
%%%%%%%%%%%%%%%%%%%%%%%%%%%%%

Several comments are in order. First, the two different initial
conditions yield very good fits to $F_2$-data, with $\chi^2/{\rm
  d.o.f.}\sim 1$, and almost identical results for $F_L$.  As remarked
in the previous Section the main difference between the two initial
conditions is their behavior at small $r$. In principle this
difference is large, but the fact that the values of $\gamma$
resulting from the fit are different for the different initial
conditions, should compensate it in a limited region of $r$. We thus
conclude that the kinematical coverage of the existing experimental
data on $F_2$ (and $F_L$) is too small to allow a discrimination of
the different UV behaviors of the two employed i.c.

Second, the fits using GBW i.c. and obtained by letting $\gamma$ vary
as a free parameter, do not show an improvement with respect to those
obtained by fixing it to $\gamma=1$. On the contrary, the fits using
MV i.c. do improve by letting $\gamma$ be a free parameter, which
takes a value slightly larger than one, $\gamma=1.13$. 

Third, although the two different fits yield pretty different values
of the initial proton saturation scale, this apparent discrepancy is
due to the different functional forms for GBW and MV i.c. If we
redefine the initial saturation scale for the MV i.c. via the
condition ${\cal N}^{MV}(r=1/Q_{s0,MV}',Y=0)=1-e^{-1/4}$ (see Section
\ref{param}), we will get $Q_{s0,\,MV}^{'2}\sim 0.19$ GeV$^2$, which
is closer to the GBW result. Therefore we conclude from our study that
the saturation scale of the proton, obtained in our fit within the
dipole model (considering only three active flavors and translational
invariant initial conditions i.e. a proton with a constant profile) at
$x=0.01$, lies in the range
$$0.19 \ \ {\rm GeV}^2 < Q_{s0}^2 < 0.25 \ \ {\rm GeV}^2.$$

Fourth, the values of $\sigma_0$ obtained from the fits are very close
to each other. This supports the assumption of translational
invariance. Furthermore, the obtained values of $\sigma_0\simeq 32$ mb correspond, assuming a Gaussian form factor for the proton \cite{Marquet:2007nf}, to a diffractive exponential slope $\sigma_0/(4 \pi)\simeq 6.5$ GeV$^{-2}$ in agreement with experimental data \cite{Aktas:2006hx}, see the comments below Eq. (\ref{dm2}).

Fifth, we have checked that the quality of the fit and the values of the parameters are stable under the restriction of the data range to the region $Q^2< 50$ GeV$^2$ (which leaves 703 data points for the fit). While in principle the dipole model should be more suitable for the description of structure functions in the region of low and moderate $Q^2$, we take this stability as a signal that there is no tension in the fit with the large-$Q^2$ data.

Finally, the agreement of our predictions for $F_L(x,Q^2)$ with the
experimental data \cite{:2008tx,Collaboration:2009na} is of the same quality as other based
on fixed-order NLO and NNLO DGLAP evolution, see the comparison
in \cite{:2008tx,Collaboration:2009na}. As discussed in \cite{Thorne:2008aj}, data at
smaller $Q^2$ may offer the possibility of discriminating different
approaches.

\subsection{Predictions for future experimental programs}
\label{predictions}

Besides available experimental data, the experimental programmes at
the LHC will test \cite{Accardi:2004be,Abreu:2007kv,Dittmar:2009ii}
our understanding of the small-$x$ behavior of the nucleon
structure. There are also proposals of future lepton-hadron colliders
\cite{eic,lhec2} in which new measurements of structure functions at
smaller $x$ would be performed. Further, the physics of high-energy
cosmic rays is expected to be influenced by small-$x$ phenomena
\cite{d'Enterria:2008jk,Armesto:2007tg}. Therefore, we find it worth
to show in Fig. \ref{figpred} our predictions for $F_2$ and $F_L$ in a
broad, yet experimentally unexplored region of $x$ and $Q^2$.

%%%%%%%%%%%%%%%%%%%%%%%%%%%%%
\FIGURE{
%\begin{figure}[htp]
%\begin{center}
\centerline{\includegraphics[width=13cm]{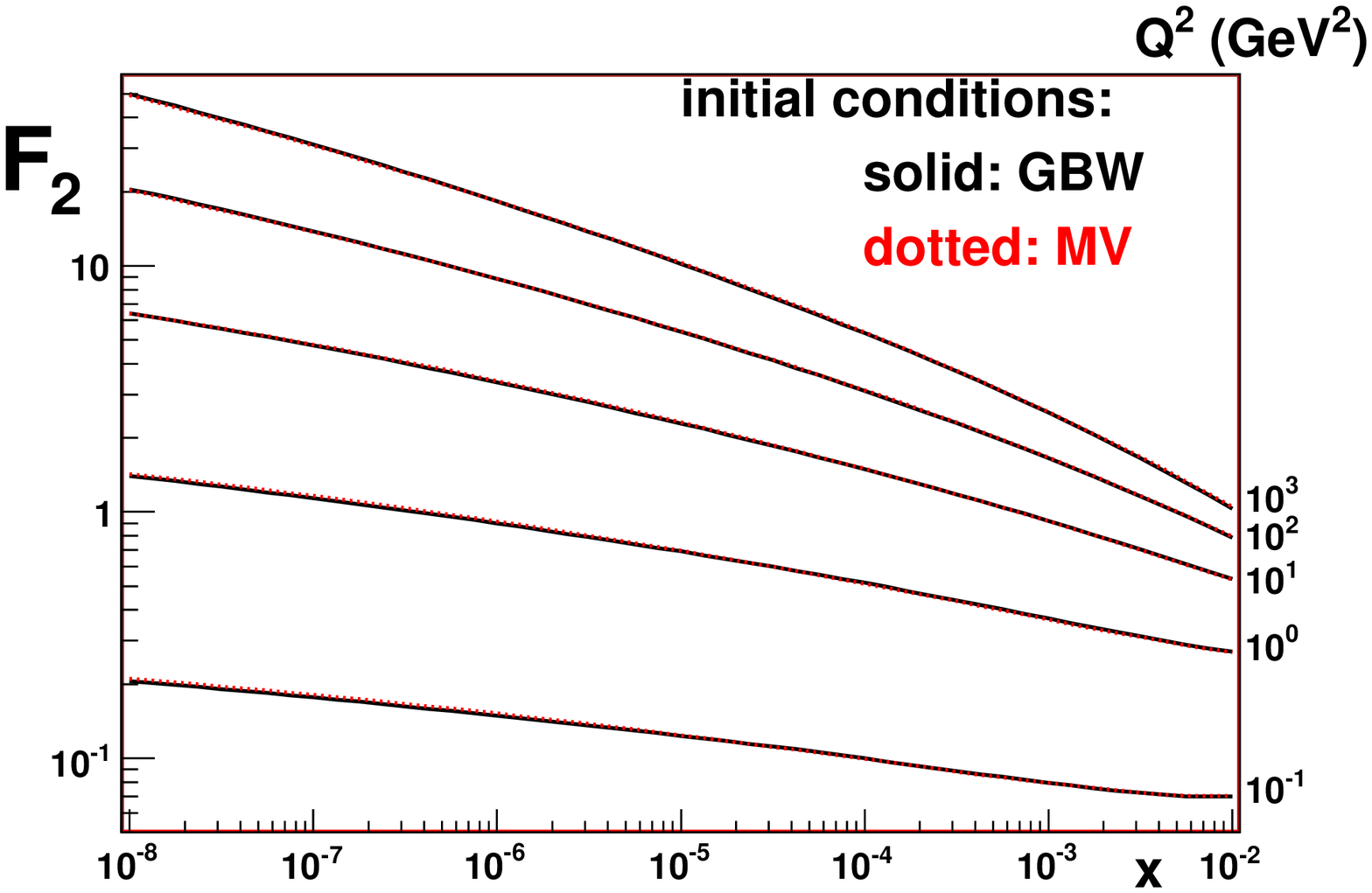}}
\centerline{\includegraphics[width=13cm]{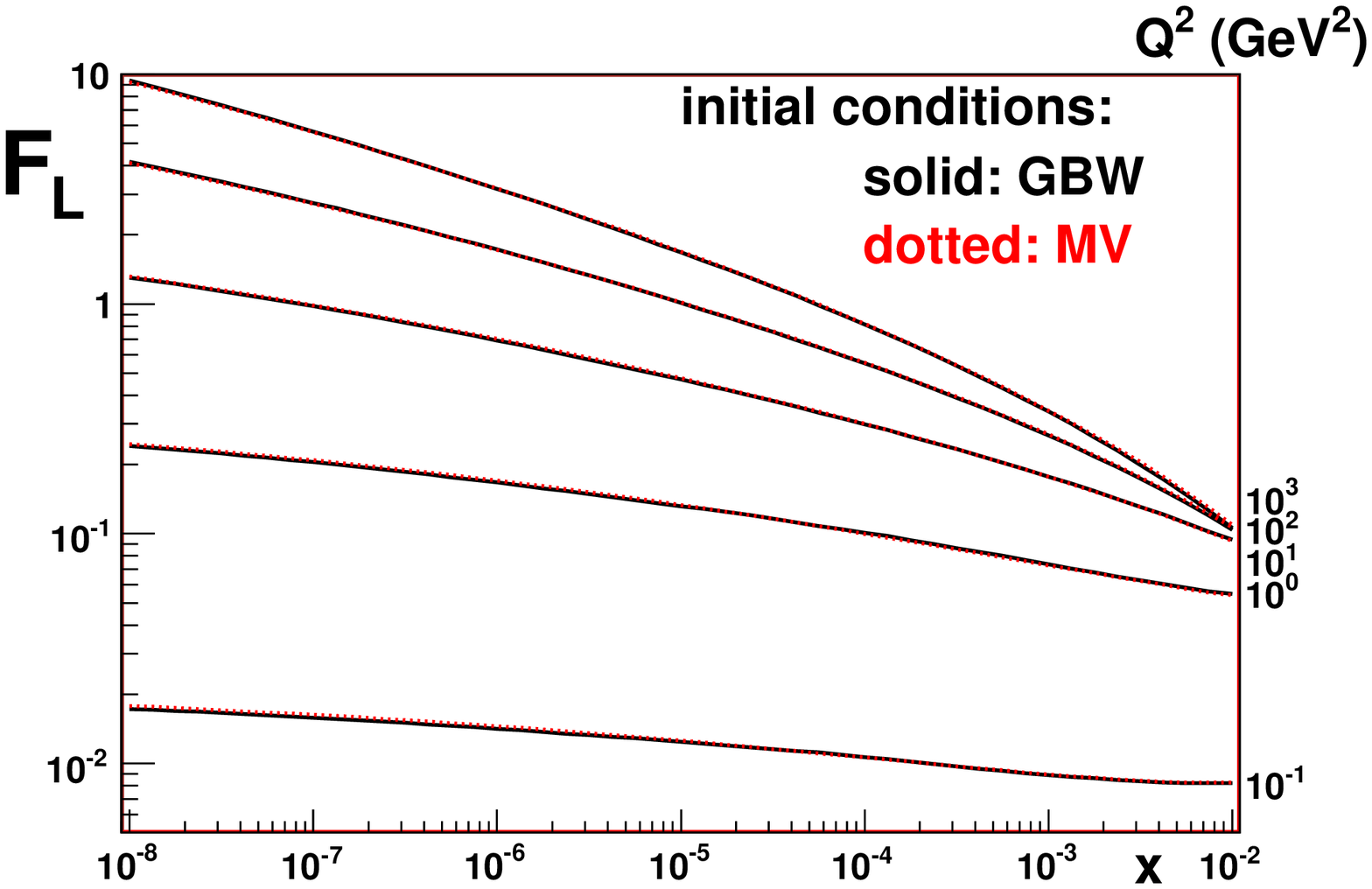}}
%\end{center}
\caption{Predictions for $F_2(x,Q^2)$ (top) and $F_L(x,Q^2)$ (bottom)
  versus $x$, for $10^{-8}\le x \le 10^{-2}$ and $Q^2=10^{-1} , 1,10,
  10^2, 10^3$ GeV$^2$ (lines from bottom to top). Solid black lines
  show the results obtained with GBW i.c., and dotted red lines those
  obtained with MV i.c.}
\label{figpred}
%\end{figure}
}
%%%%%%%%%%%%%%%%%%%%%%%%%%%%%

Two facts should be highlighted. First, the striking agreement of the
predictions --$\,$which makes them more reliable$\,$-- from both
initial conditions. Second, that at large $Q^2$ and small $x$ the
effect of saturation, namely the flattening of the structure function,
is more apparent in $F_L$ than in $F_2$. This fact stresses, in our
view, the importance of $F_L$ measurements to distinguish different
scenarios for the small-$x$ dynamics: fixed order perturbative QCD,
resummation schemes or saturation models \cite{Thorne:2008aj}.

\subsection{Parametrizations of the dipole-proton scattering
  amplitude}
\label{param}
With all the uncertainties associated to the initial condition for the
evolution fixed by the fit to $F_2$ presented in the previous
sections, we can now evolve the proton-dipole scattering amplitude to
much smaller values of $x$. Such extrapolation is completely driven by
small-$x$ evolution including running coupling corrections and can be
used to calculate several different observables relevant for the LHC
and cosmic ray physics. We have performed the evolution down to
$x=10^{-12}$. The resulting proton-dipole scattering amplitude is
plotted in Fig. \ref{pardip} for three values of $x$ ($x=10^{-2},
5\cdot10^{-6}$ and $5\cdot10^{-9}$) both for MV and GBW i.c. and
%%%% Gui
% will be 
has been 
%%%%
made public through simple fortran routines in
\cite{pweb}. From the solutions of the evolution in Fig. \ref{pardip}
we can extract the proton saturation scale $Q_{s}(x)$ through the
condition
\begin{equation}
\mathcal{N}(r=1/Q_s(x),x)=\kappa\sim\mathcal{O}(1)\,.
\label{qsdef}
\end{equation}
It is important to note that the values of $Q_{s}(x)$ presented in
Fig. \ref{qs} are dependent on the choice of $\kappa$ in
\eq{qsdef}. Following the original GBW prescription we take
\begin{equation}
  \kappa=1-\exp\left[-1/4\right]\sim 0.22\,.
\label{kap}
\end{equation}
Different choices of $\kappa$ can affect the numerical value of
$Q_{s}(x)$ by a factor $\sim 2\div3$. Keeping in mind such ambiguity
in its extraction from the numerical solutions of the evolution
equation, we can estimate the value of the proton saturation scale at
LHC energies. Using $2\to1$ kinematics to compute the smallest value
of Bjorken-$x$ probed in proton-proton collisions, $x=
(2\,M/\sqrt{s})e^{-y}$, where $M$ is the invariant mass of the
produced system (one hadron, dileptons,...), $\sqrt{s}=14$ TeV is the
collision energy and $y$ the rapidity of the produced particle, we get
(fixing $M=1$ GeV) that the saturation scale of the backward-moving
proton at the LHC at rapidities $y=0,3$ and 6 is $Q_{s}^2\simeq
0.55\div0.7$, $ 1.3\div 1.7$ and $3\div4$ GeV$^2$ respectively. Such
values are large enough to suggest that saturation effects in
proton-proton collisions at the LHC may be detectable, specially at
forward rapidities.
%%%%%%%%%%%%%%%%%%%%%%%%%%%%%
\FIGURE{
%\begin{figure}[ht]
%\begin{center}
\includegraphics[height=9cm]{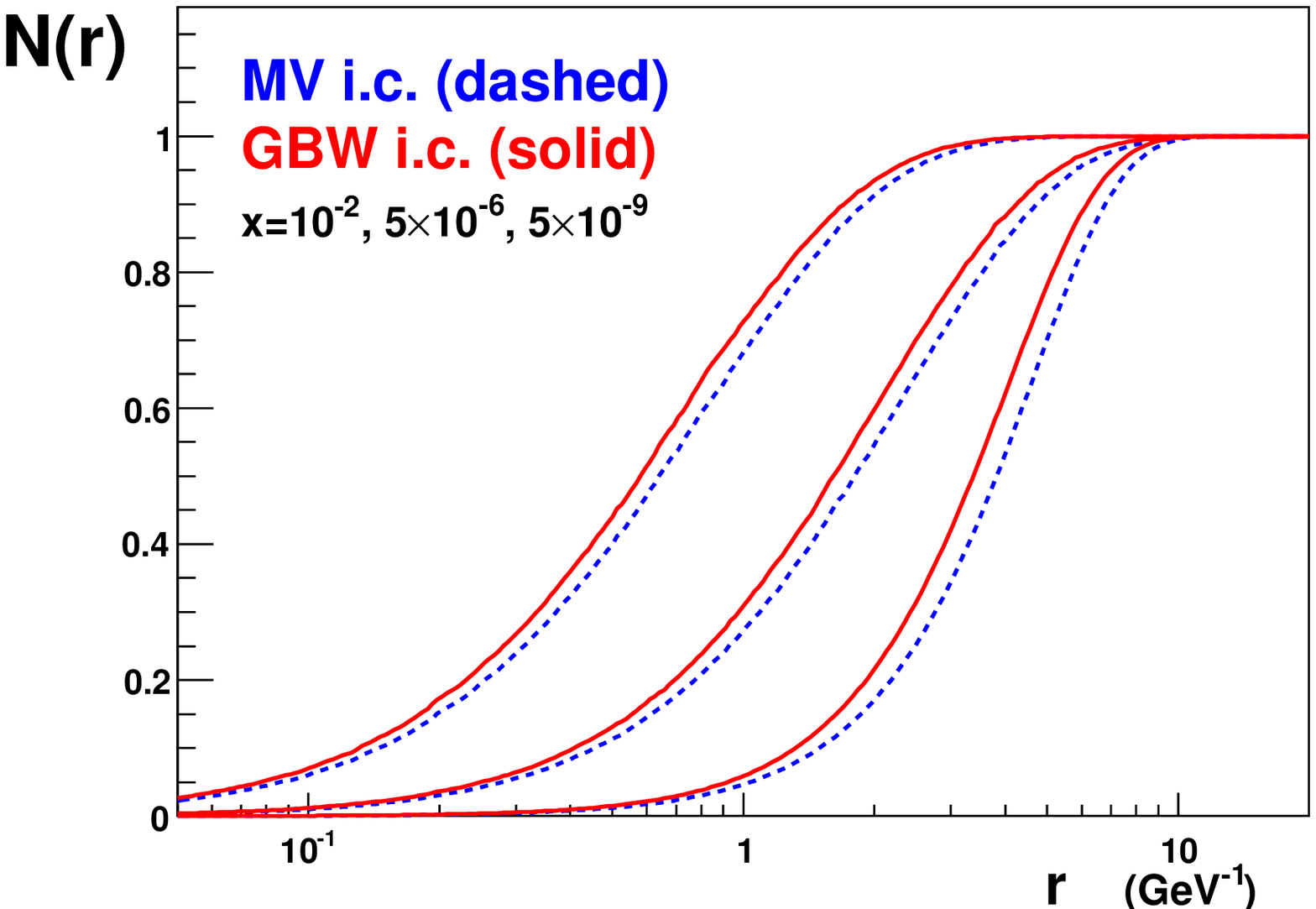}
%\end{center}
\caption{Dipole scattering amplitude obtained from the fits for
  the two different initial conditions, MV (dashed blue) and GBW
  (solid red)
  at $x=10^{-2}$, $5\cdot10^{-6}$ and $5\cdot10^{-9}$ (from right to
  left).}
\label{pardip}
}

\FIGURE{
%\begin{figure}[ht]
%\begin{center}
\includegraphics[height=9cm]{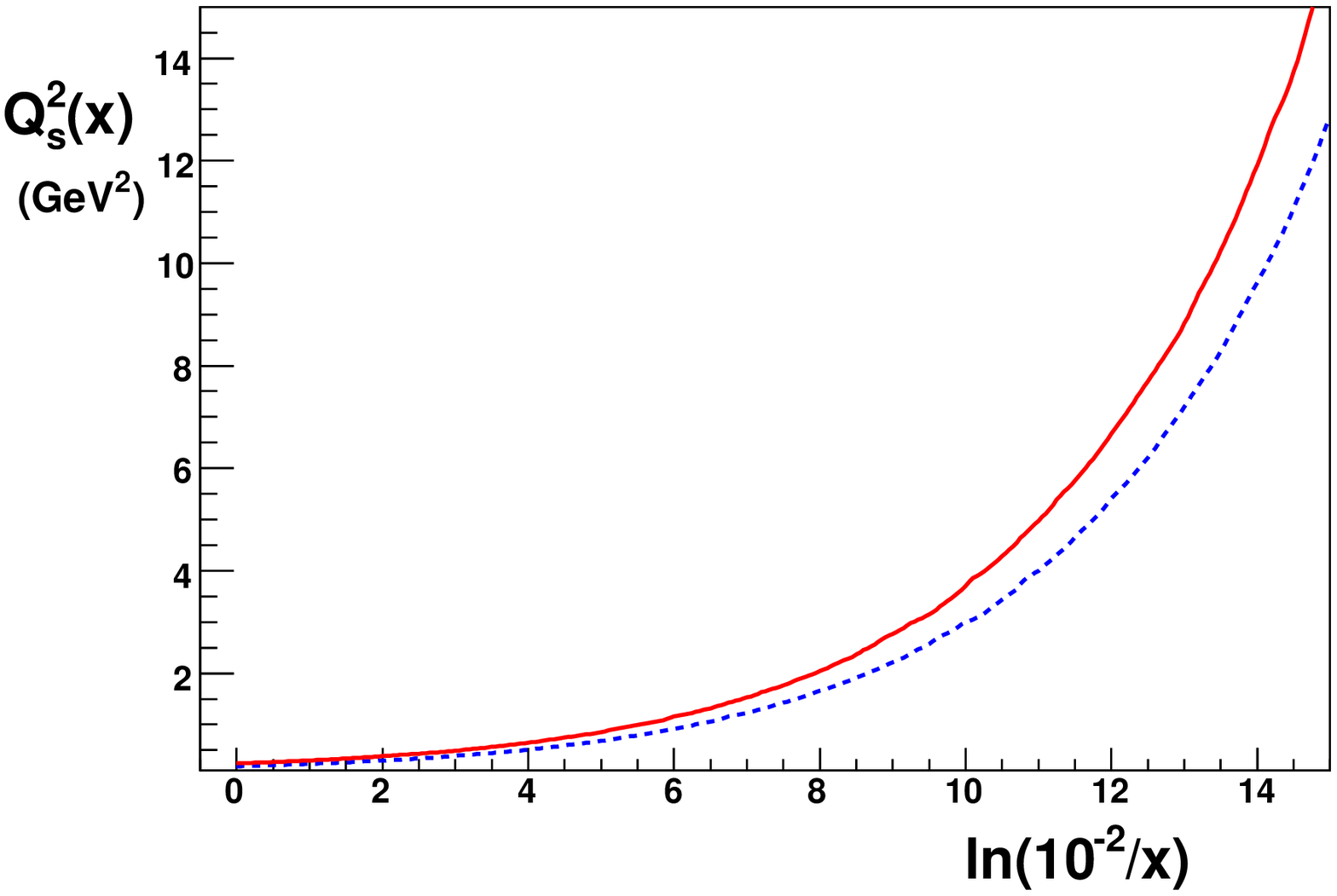}
%\end{center}
\caption{Proton saturation scale, $Q_{s}^2(x)$ versus $\ln(10^{-2}/x)$
  extracted from the solutions in Fig \ref{pardip} by the condition
  $\mathcal{N}(r=1/Q_s(x),x)=1-\exp\left[-1/4\right]$. The labeling
  follows the one in Fig. \ref{pardip}.}
\label{qs}
}
%%%%%%%%%%%%%%%%%%%%%%%%%%%%%
 
\section{Conclusions} 
\label{conclusions}

We 
%%%% Gui
%present 
presented 
%%%%
a new approach towards a systematic quantification of
parton distributions at small-$x$ directly in terms of non-linear QCD
evolution equations. This approach has become feasible thanks to the
recent calculation of the running coupling corrections to the BK
equation.  In this work we performed a global fit to the available
experimental data for $F_2(x,Q^2)$ measured in electron-proton
scattering for $x\leq10^{-2}$ and all values of $Q^2$. The calculation
of the structure functions $F_2$ and $F_L$ is done within the dipole
model under the translational invariant approximation and considering
just three active flavors. The main novelty of this work with respect
to previous phenomenological analyses is the direct use of the running
coupling BK equation to describe the small-$x$ dependence of the
structure functions. We find a very good agreement with experimental
data with only three (four) free parameters using GBW (MV) initial
conditions for the evolution. Available data on $F_L$, not included in
the fit, are also well described. We present predictions for both
$F_2$ and $F_L$ in the kinematic regime relevant for future
accelerators and ultra high-energy cosmic rays. We also provide the
evolved proton-dipole scattering amplitude down to values of
$x=10^{-12}$ through a simple computer code for public use
\cite{pweb}. Further extension of this work to nuclear targets and
hadronic and nuclear collisions is under way.

In conclusion, we find that the recent progress in our knowledge of
non-linear small-$x$ evolution brings us to an unprecedented level of
precision allowing for a direct comparison with experimental
data. This provides a solid theoretical extrapolation of parton
densities towards yet empirically unexplored kinematic
regions.

%%%%%%%%%%%%%%%%%%%%%%%%%%%%%%%%%%%%%%%%%%%%%%%%%%%%%%%%%%%%%%%%%%%%%%%%%%%%%%%

\section*{Acknowledgments} 

We would like to thank Daniele Binosi for informative and helpful
discussions, Paul Newman for information on the experimental data for
$F_L$, and M\'ario Santos and the Observational Cosmology group at
CENTRA-IST for their generosity with computing time. This work has
been supported by Ministerio de Ciencia e Innovaci\'on of Spain under
projects FPA2005-01963, FPA2008-01177 and contracts Ram\'on y Cajal
(NA and CAS); by Xunta de Galicia (Conseller\'{\i}a de Educaci\'on and
Conseller\'\i a de Innovaci\'on e Industria -- Programa Incite) (NA
and CAS); by the Spanish Consolider-Ingenio 2010 Programme CPAN
(CSD2007-00042) (NA and CAS); by the European Commission grant
PERG02-GA-2007-224770 (CAS);
and by Funda\c c\~ao para a Ci\^encia e a Tecnologia
of Portugal under project CERN/FP/83593/2008 and contract CIENCIA 2007
(JGM).

%%%%%%%%%%%%%%%%%%%%%%%%%%%%%%%%%%%%%%%%%%%%%%%%%%%%%%%%%%%%%%%%%%%%%%%%%%%%%%

\providecommand{\href}[2]{#2}\begingroup\raggedright\endgroup

%\bibliographystyle{JHEP}
%\bibliography{references}                   
\end{document}